\documentclass[conference]{IEEEtran}

\usepackage{cite}      % Written by Donald Arseneau
\usepackage{graphicx}  % Written by David Carlisle and Sebastian Rahtz
\usepackage{subfigure} % Written by Steven Douglas Cochran
\usepackage{url}       % Written by Donald Arseneau
\usepackage{stfloats}  % Written by Sigitas Tolusis
\usepackage{amsmath}   % From the American Mathematical Society
\usepackage{amssymb}   % From the American Mathematical Society

\usepackage{amsmath}
\usepackage{amsthm}
\usepackage{amssymb}
\usepackage{cite}
\usepackage{graphicx}
\usepackage{booktabs}
\usepackage{subfigure}
\newtheorem{theo}{Theorem}

\begin{document}

\title{On the Capacity of the $K$-User Cyclic Gaussian Interference Channel}

\author{Lei Zhou and Wei Yu \\
       Department of Electrical and Computer Engineering, \\
       University of Toronto, Toronto, Ontario M5S 3G4, Canada \\
       emails: \{zhoulei, weiyu\}@comm.utoronto.ca
}

\maketitle

\begin{abstract}
This paper studies the capacity region of a $K$-user cyclic Gaussian
interference channel, where the $k$th user interferes with only the
$(k-1)$th user (mod $K$) in the network. Inspired by the work of
Etkin, Tse and Wang, which derived a capacity region outer bound for the
two-user Gaussian interference channel and proved that a simple
Han-Kobayashi power splitting scheme can achieve to within one bit of the
capacity region for all values of channel parameters, this paper shows
that a similar strategy also achieves the capacity region for the
$K$-user cyclic interference channel to within a constant gap in the
weak interference regime. Specifically, a compact representation of the Han-Kobayashi achievable rate region using Fourier-Motzkin elimination is first derived, a capacity region outer bound is then established. It is
shown that the Etkin-Tse-Wang power splitting strategy gives a
constant gap of at most two bits (or one bit per dimension) in the weak interference regime. Finally, the
capacity result of the $K$-user cyclic Gaussian interference channel
in the strong interference regime is also given.
\end{abstract}

\section{Introduction}
The interference channel models a communication scenario where several
mutually interfering transmitter-receiver pairs share the same
physical medium. The interference channel is a useful model for
practical wireless network. The capacity region of the interference
channel, however, has not been completely characterized, even for the
two-user Gaussian case.

The largest achievable rate region for the two-user interference channel is due to a Han-Kobayashi strategy \cite{HK1981}, where each transmitter splits its transmit signal into a common
and a private part. The achievable rate region is the convex hull of
the union of achievable rates where each receiver decodes the common
messages from both transmitters plus the private message intended for
itself.
% using a fairly
%complex characterization involving both common-private power splitting
%and time-sharing techniques.
Recently, Chong et al.\ \cite{Chong2006} obtained an equivalent
achievable rate region but in a simpler form
by applying the Fourier-Motzkin algorithm together with a time-sharing
technique to the Han and Kobayashi's original rate region.
The optimality of the Han-Kobayashi region for the two-user Gaussian
interference channel is still an open problem in general, except in the
strong interference regime where transmission with common information
only is shown to achieve the capacity region \cite{HK1981, Carleial1978,
Sato}, and in a noisy interference regime where transmission with
private information only is shown to be sum-capacity achieving \cite{VVV,
Khandani08, Biao}.

In a recent breakthrough, Etkin, Tse and Wang \cite{Tse2007} showed that the Han-Kobayashi scheme can in fact
achieve to within one bit of the capacity region for the two-user
Gaussian interference channel for all channel parameters. Their key
insight was that the interference-to-noise ratio (INR) of
the private message should be chosen to be as close to $1$ as possible
in the Han-Kobayashi scheme. They also found a new capacity region
outer bound using a genie-aided technique.

The Etkin-Tse-Wang result applies only to the two-user interference
channel. Practical communication systems often have more than two
transmitter-receiver pairs, yet extending the one-bit result of Etkin,
Tse and Wang's work to beyond the two-user case is by no means
trivial. This is because when more than 2 users are involved, the
Han-Kobayashi private-common superposition coding strategy becomes
exceedingly complicated. It is conceivable that multiple common
messages may be needed at each transmitter, each intended to be
decoded by an arbitrary subset of receivers, thus making the
optimization of the resulting rate region difficult.  Further,
superposition coding itself may not be adequate.  Interference
alignment types of coding scheme \cite{Jafar_interference_alignment}
has been shown to be able to enlarge the achievable rate region and to
achieve to within constant gap of many-to-one and one-to-many
interference channels \cite{many-to-one}.

In the context of $K$-user Gaussian interference channels, sum-capacity results are available in the noisy interference regime
\cite{VVV, Shang_ICC2008}. Annapureddy et al.\ \cite{VVV} obtained the sum capacity for the symmetric three-user Gaussian
interference channel, the one-to-many and the many-to-one Gaussian
interference channels under the noisy interference criterion. Shang et al.\ \cite{Shang_ICC2008} studied the
fully connected $K$-user Gaussian interference channel and showed that
treating interference as noise at the receiver is sum-capacity
achieving when the transmit power and the cross channel gains are
sufficiently weak
to satisfy a certain criterion. In addition, much work has also
been carried out on the generalized degrees of freedom (as defined
in \cite{Tse2007}) of the $K$-user interference channel and its
variations \cite{Jafar_interference_alignment, Jafar_gdof_K,
Jafar_gdof_many_to_one}.

\begin{figure} [t]
\centering
\includegraphics[width=3.3in]{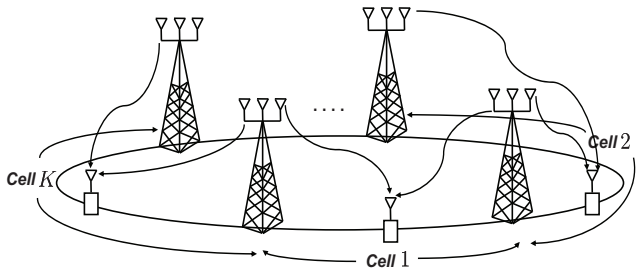}
\caption{The circular array handoff model} \label{modified_wyner_model}
\end{figure}

Instead of treating the general $K$-user interference channel, this
paper focuses on a cyclic Gaussian interference channel model, where
the $k$th user interferes with only the $(k-1)$th user. In this case,
each transmitter interferes with only one other receiver, and each
receiver suffers interference from only one other transmitter, thereby
avoiding the difficulties mentioned earlier. For the $K$-user cyclic
interference channel, the Etkin, Tse and Wang's coding strategy remains a natural one. In our previous work \cite{zhou_kuser}, we showed that such a strategy achieves the sum capacity for a symmetric channel to within two bits. The main objective of this paper is to show that this strategy also achieves to within two bits of the capacity region for the general cyclic interference channel in the weak interference regime. This paper contains an outline of the main results. Detailed proofs can be found in \cite{Lei_cyclic_jrl}.

The cyclic interference channel model is motivated by the so-called
modified Wyner model, which describes the soft handoff scenario of a
cellular network \cite{Somekh_softhandoff}. The original Wyner model
\cite{Wyner_softhandoff} assumes that all cells are arranged in a
linear array with the base-stations located at the center of each cell,
and where intercell interference comes from only the two adjacent cells.
In the modified Wyner model \cite{Somekh_softhandoff} cells are
arranged in a circular array  as shown in Fig.~\ref{modified_wyner_model}. The  mobile terminals are located along the circular array. If one assumes that the mobiles always communicate with the intended base-station to its left (or
right), while only suffering from interference due to the base-station
to its right (or left), one arrives at the $K$-user cyclic Gaussian
interference channel studied in this paper. The modified Wyner model
has been extensively studied in the literature \cite{Somekh_softhandoff, Liang_softhandoff, Sheng_softhandoff}, but often either with interference treated as noise or with the assumption of full base station cooperation. This paper studies the modified Wyner model without base station cooperation, in
which case the soft handoff problem becomes that of a cyclic interference channel.

The main results of this paper are as follows. For the $K$-user cyclic Gaussian interference channel in the weak interference regime, one can achieve to within two bits of the capacity region using the Etkin, Tse and Wang's power splitting scheme. The capacity region in the strong interference regime is also given. It is shown that transmission with common message only achieves the capacity region.

A key part of the development involves a Fourier-Motzkin elimination
procedure on the achievable rate region of the $K$-user cyclic
interference channel. To deal with the large number of inequality constraints, an induction proof needs to be used. It is shown that as compared to the two-user case, where the rate region is defined by constraints on the individual rate $R_i$, the sum rate $R_1+R_2$, and the sum rate plus an individual rate $2R_i + R_j$
($i \neq j$), the achievable rate region for the $K$-user cyclic
interference channel is defined by an additional set of constraints on
the sum rate of any arbitrary $l$ adjacent users, where $2 \le l < K$.
These four types of rate constraints completely characterize the
Han-Kobayashi region for the $K$-user cyclic interference channel.
They give rise to a total of $K^2+1$ constraints.

\section{Channel Model}

\begin{figure} [t]
\centering
\includegraphics[width=2.3in]{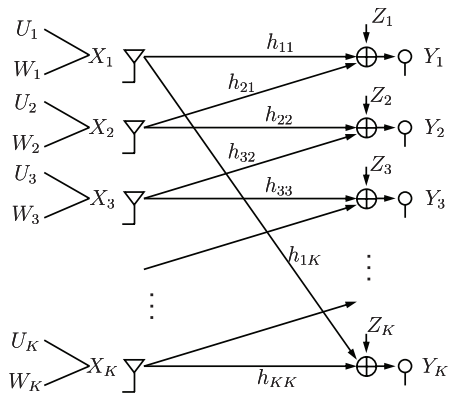}
\caption{Cyclic Gaussian interference channel} \label{cyclic_ic}
\end{figure}

The $K$-user cyclic Gaussian interference channel is first introduced
in \cite{zhou_kuser}.  It consists of $K$ transmitter-receiver pairs
as shown in Fig.~\ref{cyclic_ic}.  Each transmitter communicates with
its intended receiver while causing interference to only one neighboring
receiver. Each receiver receives a signal intended for it and an
interference signal from only one neighboring sender plus the additive
white Gaussian noise (AWGN). As shown in Fig.~\ref{cyclic_ic}, $X_1,
X_2, \cdots X_K$ and $Y_1, Y_2,
\cdots Y_K$ are complex-valued input and output signals, respectively, and $Z_i
\thicksim \mathcal{CN} (0,\sigma^2)$ is the independent and
identically distributed (i.i.d) circularly symmetric Gaussian noise at
receiver $i$. The input-output model can be written as
\begin{eqnarray}
Y_1 &=& h_{1,1}X_1 + h_{2,1}X_2 + Z_1, \nonumber \\
Y_2 &=& h_{2,2}X_2 + h_{3,2}X_3 + Z_2, \nonumber \\
& \vdots  & \nonumber \\
Y_K &=& h_{K,K}X_K + h_{1,K}X_{1} + Z_K,
\label{input_output}
\end{eqnarray}
where each $X_i$ has a power constraint $P_i$ associated with it,
i.e., $\mathbb{E}\left[|X_i|^2 \right] \le P_i$. Here, $h_{i,j}$ is the complex-valued channel gain from transmitter $i$ to receiver $j$.

The encoding-decoding procedure is described as follows. Transmitter $i$ maps a message $m_i\in\{1, 2, \cdots, 2^{nR_i} \}$ into a length $n$ codeword $X_i^n$ that belongs to a codebook $\mathcal{C}_i^n$, i.e. $X_i^n = f_i^n(m_i)$, where $f_i^n(.)$ represents the encoding function of user $i$, $i=1, 2, \cdots, K$. Codeword $X_i^n$ is then sent over a block of $n$ time instances. From the received sequence $Y_i^n$, receiver $i$ obtains an estimate $\hat{m}_i$ of the transmit message $m_i$ using a decoding function $g_i^n(.)$, i.e. $\hat{m}_i = g_i^n(Y_i^n)$. The average probability of error is defined as $P_e^n = \mathbb{E}\left[\textrm{Pr}(\cup (\hat{m}_i \neq m_i)) \right]$. A rate tuple $(R_1, R_2, \cdots, R_K)$ is said to be achievable if for an $\epsilon > 0$, there exists a family of codebooks $\mathcal{C}_i^n$, encoding functions $f_i^n(.)$, and decoding functions $g_i^n(.)$, $i=1, 2, \cdots, K$, such that $P_e^n < \epsilon$ for a sufficiently large $n$. The capacity region is the collection of  all achievable rate tuples.

Define the signal-to-noise and
interference-to-noise ratios for each user as follows\footnote{Note
that the definition of $\mathsf{INR}$ is slightly different from that
of Etkin, Tse and Wang \cite{Tse2007}.}:
\begin{equation} \label{snr_inr_def}
\mathsf{SNR}_i = \frac{|h_{i,i}|^2P_i}{\sigma^2},  \quad
\mathsf{INR}_i = \frac{|h_{i,i-1}|^2P_{i}}{\sigma^2}, \; i=1, 2, \cdots, K.
\end{equation}
The $K$-user cyclic Gaussian interference channel is said to be in the
weak interference regime if
\begin{equation} \label{weak_regime_def}
\mathsf{INR}_i \le \mathsf{SNR}_i, \quad \forall i=1, 2, \cdots, K.
\end{equation}
and the strong interference regime if
\begin{equation} \label{strong_regime_def}
\mathsf{INR}_i \ge \mathsf{SNR}_i, \quad \forall i=1, 2, \cdots, K.
\end{equation}
Otherwise, it is said to be in the mixed interference regime, which has $2^K -2$ possible combinations.

Throughout this paper, modulo arithmetic is implicitly used on the
user indices, e.g., $K+1=1$ and $1-1=K$. Note that when $K=2$, the
cyclic channel reduces to the conventional two-user interference
channel.

\section{Within Two Bits of the Capacity Region in the Weak Interference Regime}
In the two-user case, the shape of the Han-Kobayashi achievable rate region
is the union of polyhedrons (each corresponding to a fixed input
distribution) with boundaries defined by rate constraints on $R_1$,
$R_2$, $R_1+R_2$, and on $2R_1+R_2$ and $2R_2 + R_1$, respectively.
To extend Etkin, Tse and Wang's result to the general case, one needs
to find a similar rate region characterization for
the general $K$-user cyclic interference channel first.

A key feature of the cyclic Gaussian interference channel model is
that each transmitter sends signal to its intended receiver while
causing interference to {\em only one} of its neighboring receivers;
meanwhile, each receiver receives the intended signal plus the
interfering signal from {\em only one} of its neighboring transmitters.
Using this fact and with the help of Fourier-Motzkin elimination
algorithm, we show in this section that the achievable rate region of the
$K$-user cyclic Gaussian interference channel is the union of
polyhedrons with boundaries defined by rate constraints on the
individual rates $R_i$, the sum rate $R_{sum}$, the sum rate plus an
individual rate $R_{sum} + R_i$ ($i=1,2,\cdots, K$), and the sum rate
for arbitrary $l$ adjacent users ($2 \le l < K$). This last rate
constraint on arbitrary $l$ adjacent users' rates is new as compared
with the two-user case.

The preceding characterization together with outer bounds to be proved
later in the section allow us to show that the capacity region of the
$K$-user cyclic Gaussian interference channel can be achieved to
within a constant gap using the Etkin, Tse and Wang's power-splitting strategy in the weak interference regime.  However, instead of the one-bit result as obtained for the two-user interference channel \cite{Tse2007}, this section shows that without time-sharing, one can achieve to within two bits of the capacity region for the $K$-user cyclic Gaussian interference channel in the weak interference regime. The strong interference regime is treated in the next section.

\subsection{Achievable Rate Region}
\begin{theo} \label{achievable_theo}
Let $\mathcal{P}$ denote the set of probability distributions $P(\cdot)$ that factor as
\begin{eqnarray}
\lefteqn{P(q, w_1,x_1, w_2, x_2, \cdots, w_K, x_K) =}  \nonumber \\
&& \quad \quad \quad  p(q)p(x_1, w_1|q)p(x_2, w_2|q) \cdots p(x_K, w_K|q).
\end{eqnarray}
For a fixed $P \in \mathcal{P}$, let $\mathcal{R}_{\mathrm{HK}}^{(K)}(P)$ be the set of all rate tuples $(R_1, R_2, \cdots, R_K)$ satisfying
\begin{eqnarray}
0 \le R_i &\le& \min\{d_i, a_i + e_{i-1} \}, \label{achievable_Ri} \\
\sum_{j=m}^{m+l-1}R_j &\le& \min \left\{ g_m + \sum_{j=m+1}^{m+l-2}e_j + a_{m+l-1},  \right. \nonumber \\
&& \left. \quad \quad \quad \sum_{j=m-1}^{m+l-2}e_j + a_{m+l-1}\right\}, \label{achievable_Rml} \\
R_{sum} = \sum_{j=1}^{K}R_j &\le& \min \left\{ \sum_{j=1}^{K}e_j, r_1, r_2, \cdots, r_K \right\}, \label{achievable_Rsum} \\
\sum_{j=1}^{K}R_j + R_i &\le& a_i + g_i +\sum_{j=1, j \neq i }^{K} e_j ,\label{achievable_Rsi}
\end{eqnarray}
where $a_i, d_i, e_i, g_i$ and $r_i$ are defined as follows:
\begin{eqnarray}
a_i & = & I(Y_i; X_i|W_i,W_{i+1}, Q), \\
d_i & = & I(Y_i; X_i| W_{i+1}, Q), \\
e_i & = & I(Y_i; W_{i+1}, X_i|W_i, Q), \\
g_i & = & I(Y_i; W_{i+1}, X_i| Q),
\end{eqnarray}
\begin{equation}
r_i = a_{i-1} + g_i +\sum_{j=1, j \neq i, i-1}^{K} e_j,
\end{equation}
and the range of indices are $i, m = 1, 2, \cdots, K$ in
(\ref{achievable_Ri}) and (\ref{achievable_Rsi}), $l = 2, 3, \cdots,
K-1$ in (\ref{achievable_Rml}).
Define
\begin{equation}
\mathcal{R}_{\mathrm{HK}}^{(K)} = \bigcup_{P \in \mathcal{P}} \mathcal{R}_{\mathrm{HK}}^{(K)}(P).
\end{equation}
Then $\mathcal{R}_{\mathrm{HK}}^{(K)}$ is an achievable rate region for the $K$-user cyclic interference channel.

\end{theo}

\begin{IEEEproof}
The achievable rate region can be proved by the Fourier-Motzkin
algorithm together with an induction step. The proof follows the
Kobayashi and Han's strategy \cite{HK2007}
of eliminating a common message at each step.
Details are available in \cite{Lei_cyclic_jrl}.
\end{IEEEproof}

In the above achievable rate region, (\ref{achievable_Ri}) is the
constraint on the achievable rate of an individual user,
(\ref{achievable_Rml}) is the constraint on the achievable sum rate
for any $l$ adjacent users ($2 \le l < K$), (\ref{achievable_Rsum}) is
the constraint on the achievable sum rate of all $K$ users, and
(\ref{achievable_Rsi}) is the constraint on the achievable sum rate
for all $K$ users plus a repeated user. %We can also think of
%(\ref{achievable_Ri}) to (\ref{achievable_Rsi}) as the sum-rate
%constraints for arbitrary $l$ adjacent users, where $l=1$ for
%(\ref{achievable_Ri}), $2 \le l <K$ for (\ref{achievable_Rml}), $l=K$
%for (\ref{achievable_Rsum}) and $l=K+1$ for (\ref{achievable_Rsi}).

From (\ref{achievable_Ri}) to (\ref{achievable_Rsi}), there are a
total of $K + K(K-2) + 1 + K =K^2 +1$ constraints. Together they
describe the shape of the achievable rate region under a fixed input
distribution.  The quadratic growth in the number of constraints as a
function of $K$ makes the Fourier-Motzkin elimination of the
Han-Kobayashi region quite complex. An induction needs to be used to deal with the large number of the constraints.

As an example, for the two-user Gaussian interference channel, there
are $2^2+1 = 5$ rate constraints, corresponding to that of $R_1$,
$R_2$, $R_1 +R_2$, $2R_1 +R_2$ and $2R_2 +R_1$, as in \cite{HK1981,
HK2007, Chong2006, Tse2007}. Specifically, substituting $K=2$ in
Theorem~\ref{achievable_theo} gives us the following achievable rate
region:
\begin{eqnarray}
0 \le R_1 &\le& \min \{ d_1, a_1 + e_2\},  \label{R1_constraint}\\
0 \le R_2 &\le& \min \{ d_2, a_2 + e_1\},  \label{R2_constraint}\\
R_{1}+R_{2} &\le& \min\{e_1 +e_2, a_1 + g_2, a_2 + g_1  \}, \\
2R_1 + R_2 &\le& a_1 + g_1 + e_2, \\
2R_2 + R_1 &\le& a_2 + g_2 + e_1, \label{R2_lastone}
%-R_1 &\le& 0, \\
%-R_2 &\le& 0 \label{R2_lastone}.
\end{eqnarray}
which is exactly the Theorem~D of \cite{HK2007}.

\subsection{Capacity Region Outer Bound}
\begin{theo} \label{outerbound_theo}
For the $K$-user cyclic  Gaussian interference channel in the weak
interference regime, the capacity region is included in the following set of rate tuples $(R_1, R_2, \cdots, R_K)$:
\begin{eqnarray}
R_i &\le& \lambda_i  \label{outer_Ri}, \\
\sum_{j=m}^{m+l-1}R_j &\le& \min \left\{ \gamma_m + \sum_{j=m+1}^{m+l-2}\alpha_j + \beta_{m+l-1}, \right. \nonumber \\
&&\left. \quad \quad \mu_m + \sum_{j=m}^{m+l-2}\alpha_j + \beta_{m+l-1}\right\}, \label{outer_Rml}\\
\sum_{j=1}^{K}R_j &\le& \min \left\{ \sum_{j=1}^{K}\alpha_j, \rho_1, \rho_2, \cdots, \rho_K \right\}, \label{outer_Rsum} \\
\sum_{j=1}^{K}R_j + R_i &\le& \beta_i + \gamma_i +\sum_{j=1, j \neq i }^{K} \alpha_j, \label{outer_Rsi}
\end{eqnarray}
where the ranges of the indices $i$, $m$, $l$ are as defined in
Theorem~\ref{achievable_theo}, and
\begin{eqnarray}
\alpha_i &=& \log \left(1+ \mathsf{INR}_{i+1} + \frac{\mathsf{SNR}_i}{1+\mathsf{INR}_i} \right),\label{alpha_i} \\
\beta_i &=& \log \left( \frac{1+\mathsf{SNR}_i}{1+\mathsf{INR}_i} \right), \\
\gamma_i &=& \log \left( 1 + \mathsf{INR}_{i+1} + \mathsf{SNR}_i \right), \\
\lambda_i &=& \log (1 + \mathsf{SNR}_i), \\
\mu_i &=& \log (1+ \mathsf{INR}_i), \\
\rho_i &=&\beta_{i-1} + \gamma_i + \sum_{j=1, j\neq i, i-1}^{K} \alpha_j. \label{rho_i}
\end{eqnarray}
\end{theo}

\begin{IEEEproof}
Genie-aided bounding techniques are used to prove the theorem. See \cite{Lei_cyclic_jrl} for details.
\end{IEEEproof}

\subsection{Capacity Region to Within Two Bits}
\begin{theo} \label{twobit_theo}
For the $K$-user cyclic Gaussian interference channel in the weak
interference regime, the fixed Etkin, Tse and Wang's power-splitting strategy
achieves to within two bits of the capacity
region\footnote{If a rate pair $(R_1, R_2, \cdots, R_K)$ is achievable and $(R_1+k, R_2+k, \cdots, R_K+k)$ is outside the capacity region, then $(R_1, R_2, \cdots, R_K)$ is said to be within $k$ bits of the capacity region.}.
\end{theo}

\begin{IEEEproof}
Applying the Etkin, Tse and Wang's power-splitting strategy (i.e., $\mathsf{INR}_{ip} =
\min (\mathsf{INR}_i, 1)$) to Theorem~\ref{achievable_theo},
parameters $a_i, d_i, e_i, g_i$ can be easily calculated as follows:
\begin{eqnarray}
a_i &=&  \log \left( 2 + \mathsf{SNR}_{ip}  \right) - 1, \label{a_i}\\
d_i &=&  \log \left( 2 + \mathsf{SNR}_i  \right) - 1, \label{d_i_2} \\
e_i &=&  \log \left( 1 + \mathsf{INR}_{i+1} + \mathsf{SNR}_{ip} \right) - 1, \\
g_i &=& \log \left( 1 + \mathsf{INR}_{i+1} + \mathsf{SNR}_i \right) - 1. \label{g_i}
\end{eqnarray}

To prove that the achievable rate region described by the above $a_i, d_i, e_i, g_i$ is within two bits of the outer
bound in Theorem~\ref{outerbound_theo}, we need to show that each of the rate constraints in  (\ref{achievable_Ri})-(\ref{achievable_Rsi}) is within
two bits of their corresponding outer bound in (\ref{outer_Ri})-(\ref{outer_Rsi}),
i.e., the following inequalities hold for all $i$, $m$, $l$ in the
ranges defined in Theorem~\ref{achievable_theo}:
\begin{eqnarray}
\delta_{R_i} &<& 2, \label{delta_Ri} \\
\delta_{R_m+\cdots+R_{m+l-1}} &<& 2l, \\
\delta_{R_{sum}} &<& 2K, \\
\delta_{R_{sum}+R_i} &<& 2(K+1), \label{delta_RsumRi}
\end{eqnarray}
where $\delta_{(\cdot)}$ is the difference between the achievable rate
in Theorem~\ref{achievable_theo} and its corresponding outer bound in
Theorem ~\ref{outerbound_theo}. A complete proof can be found in \cite{Lei_cyclic_jrl} .
\end{IEEEproof}

\section{Capacity Region in the Strong Interference Regime}

The results so far in the paper pertain only to the weak interference regime, where $\mathsf{SNR}_i \ge \mathsf{INR}_i$, $\forall i$. In the strong interference regime, where $\mathsf{SNR}_i \le \mathsf{INR}_i$, $\forall i$, the capacity result in \cite{HK1981} \cite{Sato} for the two-user Gaussian interference channel can be easily extended to the $K$-user  cyclic case.

\begin{theo} \label{strong_theo}
For the $K$-user cyclic Gaussian interference channel in the strong
interference regime, the capacity region is given by the set of
$(R_1,R_2,\cdots,R_K)$ such that
\begin{equation} \label{capacity_strong}
\left\{
  \begin{array}{l}
R_i \le \log (1 + \mathsf{SNR}_i) \\
R_i + R_{i+1} \le \log (1 + \mathsf{SNR}_i + \mathsf{INR}_{i+1}),
  \end{array}
\right.
\end{equation}
for $i=1, 2, \cdots, K$. In the very strong interference
regime where $\mathsf{INR}_i \ge (1 +  \mathsf{SNR}_{i-1})\mathsf{SNR}_i,
\forall i$, the capacity region is the set of $(R_1,R_2,\cdots,R_K)$ with
\begin{equation} \label{capacity_very_strong}
R_i \le \log (1 + \mathsf{SNR}_i), \;\; i=1, 2, \cdots, K.
\end{equation}
\end{theo}

\begin{IEEEproof}
{\em{Achievability}}: It is easy to see that (\ref{capacity_strong})
is in fact the intersection of the capacity regions of
$K$ multiple-access channels:
\begin{equation}
\label{capacityregion_mac}
\bigcap_{i=1}^{K}\left\{
(R_i,R_{i+1}) \left|  \begin{array}{l}
R_i \le \log (1 + \mathsf{SNR}_i) \\
R_{i+1} \le \log (1 + \mathsf{INR}_{i+1}) \\
R_i + R_{i+1} \le \log (1 + \mathsf{SNR}_i + \mathsf{INR}_{i+1}).
  \end{array}
\right.  \right \}.
\end{equation}
Each of these regions corresponds to that of a multiple-access channel with $W_i^n$ and $W_{i+1}^n$ as inputs and $Y_i^n$ as output (with $U_i^n=U_{i+1}^n=\emptyset$).  Therefore, the rate region (\ref{capacity_strong}) can be achieved by setting  all the input signals to be common messages.  This completes the achievability part.

{\em{Converse}}: The converse proof follows the idea of \cite{Sato}. The key ingredient is to show that for a genie-aided Gaussian interference channel to be defined later, in the strong interference regime, whenever a rate tuple $(R_1, R_2, \cdots, R_K)$ is achievable, i.e., $X_i^n$ is decodable at receiver $i$, $X_i^n$ must also be decodable at $Y_{i-1}^n$,  $i=1,2,\cdots, K$.

The genie-aided Gaussian interference channel is defined by the Gaussian interference channel (see Fig.~\ref{cyclic_ic}) with genie $X_{i+2}^n$ given to receiver $i$. The capacity region of the $K$-user cyclic Gaussian interference channel must be resided inside the capacity region of the genie-aided one.

%First, the reliable recoding of $X_i^n$ at receiver $i$ requires
%\begin{equation} \label{decoding_requirement}
%R_i \le \log (1 + \mathsf{SNR}_i)
%\end{equation}

Assume that a rate tuple $(R_1, R_2, \cdots, R_K)$ is achievable for the $K$-user cyclic Gaussian interference channel. In this case, after $X_i^n$ is decoded, with the knowledge of the genie $X_{i+2}^n$, receiver $i$ can construct the following signal:
\begin{eqnarray}
\widetilde{Y}_i^n &=& \frac{h_{i+1,i+1}}{h_{i+1,i}}(Y_i^n - h_{i,i}X_i^n) + h_{i+2, i+1}X_{i+2}^n \nonumber \\
&=& h_{i+1, i+1}X_{i+1}^n + h_{i+2, i+1}X_{i+2}^n + \frac{h_{i+1,i+1}}{h_{i+1,i}}Z_{i}^n, \nonumber
\end{eqnarray}
which contains the signal component of $Y_{i+1}^n$ but with less noise since $|h_{i+1,i}| \ge |h_{i+1,i+1}|$ in the strong interference regime. Now, since $X_{i+1}^n$ is decodable at receiver $i+1$, it must also be decodable at receiver $i$ using the constructed $\widetilde{Y}_i^n$. Therefore, $X_i^n$ and
$X_{i+1}^n$ are both decodable at receiver $i$. As a result, the achievable rate region of $(R_i, R_{i+1})$ is bounded by the capacity region of the multiple-access channel $(X_i^n, X_{i+1}^n, Y_i^n)$, which is shown in (\ref{capacityregion_mac}). Since (\ref{capacityregion_mac}) reduces to (\ref{capacity_strong}) in the strong interference regime, we have shown that (\ref{capacity_strong}) is an outer bound of the $K$-user
cyclic Gaussian interference channel in the strong  interference regime. This completes the converse proof.

In the very strong interference regime where $\mathsf{INR}_i \ge (1 +  \mathsf{SNR}_{i-1})\mathsf{SNR}_i, \forall i$, it is easy to verify that the second constraint in (\ref{capacity_strong}) is no longer active. This results in the capacity region (\ref{capacity_very_strong}).
\end{IEEEproof}

\section{Concluding Remarks}
This paper studies the capacities and the coding strategies for the
$K$-user cyclic Gaussian interference channel in the weak and the strong interference regimes. An achievable rate region based on the Han-Kobayashi power splitting strategy is first derived; a corresponding capacity region outer bound is then obtained using genie-aided bounding techniques. This paper shows that in the weak interference regime, the Etkin, Tse and Wang's power-splitting strategy achieves to within two bits of the capacity region. The capacity result for the $K$-user cyclic Gaussian interference channel in the strong interference regime is a straightforward extension of the corresponding two-user case. However, in the mixed interference regime, although the constant gap result may well continue to hold, the proof becomes considerably more complicated, as different mixed scenarios need to be enumerated and the corresponding outer bounds derived.

\bibliographystyle{IEEEtran}
\bibliography{IEEEabrv,./ref/main}

\end{document}